\documentclass[12pt]{article}
\usepackage{makeidx}
\usepackage{amsmath}
\usepackage{amsfonts}
\usepackage{amssymb}
\usepackage{epsfig}
\usepackage[cp1251]{inputenc}
\usepackage{graphicx}

\def\bra #1{\langle #1\vert}
\def\ket #1{\vert #1\rangle}

\newcounter{defin}
\newcounter{lemma}
\newcounter{theorem}
\newcounter{proposition}
\newcounter{example}
\newenvironment{lemma}{\par\refstepcounter{lemma}     \textbf{Lemma \thelemma.} }{\rm\par}
\newenvironment{theorem}{\par\refstepcounter{theorem}     \textbf{Theorem \thetheorem.}\ }{\rm\par}

\newcommand{\Tr}{\mathrm{Tr}}

\begin{document}

\title{Quantum information aspects of approximate position measurement}
\author{A. S. Holevo, V. I. Yashin \\
Steklov Mathematical Institute, RAS, Moscow, Russia}
\date{}
\maketitle

\begin{abstract}
We perform a quantum information analysis for multi-mode Gaussian
 approximate position measurements, underlying
 noisy homodyning in quantum optics. The
\textquotedblleft Gaussian maximizer\textquotedblright\ property is
established for the entropy reduction of these measurements which provides
explicit formulas for computations including their entanglement-assisted capacity.
The case of one mode is discussed in detail.

\end{abstract}

\section{\protect\bigskip Introduction}\label{s0}

Among quantitative characteristics of quantum measurement, the entropy
reduction and the entanglement-assisted classical information capacity
(called also measurement strength or information gain, depending on different operational
interpretations) play a significant role. We
refer a reader to \cite{winmas}, \cite{Sh-ERQM}, \cite{h5}, \cite{H1}, \cite%
{H2} and to \cite{BRW} where one can find also a detailed survey and further
references. In \cite{H3} we presented a study of these quantities for a
class of multi-mode Gaussian approximate joint
position-momentum measurements (or, in the context of quantum optics, \textquotedblleft
noisy heterodyning\textquotedblright ). In particular, for this class of
measurements we established the \textquotedblleft Gaussian
maximizer\textquotedblright\ property of the entropy reduction which allowed
for an explicit computation of this quantity and of the  related
entanglement-assisted capacity. In the present paper we perform a similar
analysis for another important class of multi-mode Gaussian measurements,
namely, of approximate position measurements underlying the
\textquotedblleft noisy homodyning\textquotedblright\ which is in a sense
opposite to the \textquotedblleft heterodyning\textquotedblright\  (see \cite%
{caves}, \cite{hall}, \cite{sera} for a detailed physical description of the relevant
measurement processes). The \textquotedblleft Gaussian
maximizer\textquotedblright\ property is established in theorem \ref{t3}
which also gives the explicit formula for computations. The
entanglement-assisted capacity is considered in sec. \ref{s3} and the case
of one mode is discussed in detail in sec. \ref{s4}.

\section{\protect\bigskip Entropy reduction of quantum observable}\label{s1}

Let $\mathcal{H}$ be the separable Hilbert space of a quantum system, and
let $\mathfrak{S}(\mathcal{H})$ be the convex set of quantum states (density
operators on $\mathcal{H}$). Let $(\mathcal{X},\mathcal{F})$ be a standard
measurable space of the outcomes of a measurement described by an \emph{%
completely positive (c.p.) instrument} $\mathcal{M=}\left\{ \mathcal{M}%
(A)[\cdot ],A\in \mathcal{F}\right\} ,$ i.e. operation-valued measure where
for each $A$ the map $\rho \rightarrow \mathcal{M}(A)[\rho ],\,\rho \in
\mathfrak{S}(\mathcal{H}),$ is c.p. trace-nonincreasing, and $\mathcal{M}(%
\mathcal{X})$ is trace-preserving (t.p.) (see, e.g. \cite{BL-2}). Let $%
M=\{M(A),A\in \mathcal{F}\}$ be the associated \emph{\ observable} i.e. a
probability operator-valued measure (POVM) on $(\mathcal{X},\mathcal{F})$
such that $M(A)=\mathcal{M}^{\ast }(A)[I],$ where $I$ \ is the identity
operator on $\mathcal{H}$. \emph{The probability distribution} of the
observable $M$ in the state $\rho \in \mathfrak{S}(\mathcal{H})$ is given by
the formula
\begin{equation*}
P_{\rho }(A)=\Tr \rho M(A),\quad A\in \mathcal{F}.
\end{equation*}%
As shown in \cite{Oz} for each $\rho \in \mathfrak{S}(\mathcal{H})$ there is
a family of \emph{posterior states} $\left\{ \hat{\rho}(x)\right\} $ such
that
\begin{equation*}
\mathcal{M}(A)\left[ \rho \right] =\int_{A}\hat{\rho}(x)P_{\rho }(dx).
\end{equation*}%
for any state $\rho \in \mathfrak{S}(\mathcal{H})$. The \emph{entropy
reduction} of the instrument is defined as%
\begin{equation*}
ER(\rho ,\mathcal{M})=\mathrm{H}(\rho )-\int_{\mathcal{X}}\,\mathrm{H}(\hat{%
\rho}(x))P_{\rho }(dx),
\end{equation*}%
where $\mathrm{H}(\rho )=-\Tr \rho \log \rho ,$ provided $\mathrm{H}%
(\rho )<\infty $ \cite{Oz}, \cite{Sh-ERQM}.

We will deal with the special class of instrument, the observables of which
have bounded \emph{operator-valued density}, more precisely,
\begin{equation}
M(A)=\int_{A}m(x)\,\mu (dx),\quad A\in \mathcal{F},  \label{den}
\end{equation}%
where $\mu $ is a $\sigma $-finite measure on the $\sigma $-algebra $%
\mathcal{F}$ such that $m(x)=V(x)^{\ast }V(x)$, where $V(x)$ is a weakly
measurable function with values in the algebra of bounded operators on $%
\mathcal{H}$ satisfying
\begin{equation*}
\int_{\mathcal{X} }V({x})^{\ast }V({x})\mu (dx)=I,
\end{equation*}%
and the integral weakly converges \cite{H1}. With any measurable
factorization of $m(x)$ one can associate an \emph{efficient instrument}
(see \cite{Oz}, \cite{Sh-ERQM})
\begin{equation*}
\mathcal{M}(A)\left[ \rho \right] =\int_{A}V({x}) \rho V({x})^{\ast }\mu
(dx).
\end{equation*}%
Then the probability distribution $P_{\rho }(dx)$ has the density
\begin{equation}
p_{\rho }(x)=\Tr \rho V({x})^{\ast }V({x})  \label{opd}
\end{equation}%
with respect to the measure $\mu ,$ while the family of posterior states is
\begin{equation}
\hat{\rho}(x)=p_{\rho }(x)^{-1}V(x)\rho V(x)^{\ast }  \label{post2}
\end{equation}

The entropy reduction of the efficient instrument is then
\begin{equation}
ER(\rho ,M)=\mathrm{H}(\rho )-\int_{\Omega }p_{\rho }(x)\,\mathrm{H}(\hat{%
\rho}(x))\mu (dx),  \label{er-1}
\end{equation}
where the posterior states are given by (\ref{post2}).
In \cite{Oz} it was shown that the entropy reduction of an efficient
instrument is nonnegative. In \cite{Sh-ERQM} the entropy reduction was
related to the quantum mutual information of the instrument and hence it is a
concave, subadditive, lower semicontinuous function of $\rho .$

An essential observation made in \cite{h5}, \cite{H3} is that $\mathrm{H}(%
\hat{\rho}(x))$ in the entropy reduction (\ref{er-1}) depend only on $m(${$x$%
}$)$ (i.e. on the observable $M$) and not on the way of its measurable
factorization (i.e. the choice of an efficient instrument), which justifies
the notation $ER(\rho ,M).$

\section{\protect\bigskip Gaussian maximizers for the entropy reduction}\label{s2}

Our framework will be a bosonic system with $s$ modes described by the
canonical position-momentum operators $q=[q_{1},\dots
,q_{s}]^{t},\,p=[p_{1},\dots ,p_{s}]^{t}$ (see e.g. \cite{QSCI}, \cite{sera}%
). It will be convenient to take the Schr\"{o}dinger (position)
representation space $\mathcal{H}=L^{2}(\mathbb{R}^{s})$ with the operators $%
q_{j}=\xi _{j},\,p_{j}=i^{-1}\frac{\partial }{\partial \xi _{j}}.$ We denote
by
\begin{equation*}
D(x,y)=\mathrm{e}^{-ix^{t}y/2}\mathrm{e}^{iy^{t}q}\mathrm{e}%
^{-ix^{t}p},\quad x,y\in \mathbb{R}^{s}
\end{equation*}%
the unitary displacement operators.

We will be interested in approximate position measurement in $s$ modes. In
quantum optics this underlies the multi-mode noisy homodyning, as measurement of any quadrature
of the radiation field can be reduced to the measurement of position by a Bogoljubov transformation.
In this case $\mathcal{X}=\mathbb{R}^{s}$ and the POVM is
\begin{eqnarray}
M(d^{s}x) &=&\exp \left[ -\frac{1}{2}\left( q-x\right) ^{t}\beta ^{-1}\left(
q-x\right) \right] \frac{d^{s}x}{\sqrt{\left( 2\pi \right) ^{s}\det \beta }}
\notag \\
&=&D(x)\exp \left[ -\frac{1}{2}q^{t}\beta ^{-1}q\right] D(x)^{\ast }\frac{%
d^{s}x}{\sqrt{\left( 2\pi \right) ^{s}\det \beta }} ,  \label{apprqs1}
\end{eqnarray}%
where $\beta $ is a real positive definite covariance matrix of
the measurement noise, and $D(x)=D(x,0)=\exp \left( -ix^{t}p\right) $ is the
position displacement operator. Notice that POVM $M$ has the form (\ref{den}%
) with the bounded operator-valued density $m(x)$, hence we are in the
situation where the formulas (\ref{post2}) and (\ref{er-1}) are applicable.

Denote
\begin{equation*}
V(x)=V^{\ast }(x)=\sqrt{m(x)}=\left[ \left( 2\pi \right) ^{s}\det \beta %
\right] ^{-1/4}\exp \left[ -\frac{1}{4}\left( q-x\right) ^{t}\beta
^{-1}\left( q-x\right) \right] .
\end{equation*}
Let $\rho $ be a state and $\rho _{u,v}=D(u,v)\rho D(u,v)^{\ast }$ a
displaced state. Then
\begin{eqnarray*}
&& V(x)\rho _{u,v}V(x)=\\ &=&D(u,v)\left( D(u,v)^{\ast }V(x)D(u,v)\right) \rho
\left( D(u,v)^{\ast }V(x)\rho D(u,v)\right) D(u,v)^{\ast } \\
&=&D(u,v)V(x+u)\rho V(x+u)D(u,v)^{\ast }.
\end{eqnarray*}%
It follows that
\begin{equation*}
p_{\rho _{u,v}}(x)=p_{\rho }(x+u),\quad \hat{\rho}_{u,v}(x)=D(u,v)\hat{\rho}%
(x+u)D(u,v)^{\ast },
\end{equation*}%
and
\begin{equation}
ER(\rho _{u,v},M)=ER(\rho ,M)  \label{zerom}
\end{equation}%
by the unitary invariance of the entropy. Therefore in what follows we can
restrict to centered states $\rho$ ($\Tr \,\rho \,q_{j}=\mathrm{%
\Tr }\,\rho \,p_{j}=0$).

Consider the quantum state $\rho _{\alpha }$ $\in \mathfrak{S}(\mathcal{H})$
which is centered Gaussian density operator in $\mathcal{H}$ with the
covariance $2s\times 2s-$matrix
\begin{equation}
\alpha =\left[
\begin{array}{cc}
\alpha _{qq} & \alpha _{qp} \\
\alpha _{pq} & \alpha _{pp}%
\end{array}%
\right] ,  \label{cm}
\end{equation}%
where
\begin{equation*}
\alpha _{qq}=\alpha _{qq}^{t}=\left[ \Tr \,\rho _{\alpha
}q_{j}q_{k}\right] _{j,k=1,\dots s},\quad \quad \alpha _{pp}=\alpha
_{pp}^{t}=\left[ \Tr \,\rho _{\alpha }p_{j}p_{k}\right]
_{j,k=1,\dots s},
\end{equation*}%
\begin{equation*}
\alpha _{qp}=\alpha _{pq}^{t}=\left[ \frac{1}{2}\Tr \,\rho
_{\alpha }\left( q_{j}p_{k}+p_{k}q_{j}\right) \right] _{j,k=1,\dots s}.
\end{equation*}%
It is explicitly given by the kernel in the Schr\"{o}dinger representation
\begin{equation}
\langle \xi |\rho _{\alpha }|\xi ^{\prime }\rangle =\frac{1}{\sqrt{\left(
2\pi \right) ^{s}\det \alpha _{qq}}}\exp \left[ -\frac{1}{2}\theta \left(
\xi ,\xi ^{\prime }\right) \right] ,  \label{mm0}
\end{equation}%
where%
\begin{eqnarray}
\theta \left( \xi ,\xi ^{\prime }\right) &=&\frac{1}{4}\left( \xi ^{\prime
}+\xi \right) ^{t}\alpha _{qq}^{-1}\left( \xi ^{\prime }+\xi \right) +\left(
\xi ^{\prime }-\xi \right) ^{t}\left( \alpha _{pp}-\alpha _{pq}\alpha
_{qq}^{-1}\alpha _{qp}\right) \left( \xi ^{\prime }-\xi \right)  \notag \\
&+&i\left( \xi ^{\prime }+\xi \right) ^{t}\alpha _{qq}^{-1}\alpha
_{qp}\left( \xi ^{\prime }-\xi \right) ,  \label{theta}
\end{eqnarray}%
see Appendix.

Let%
\begin{equation*}
\rho _{\alpha }(x)=V(x)\rho _{\alpha }V(x),
\end{equation*}%
then the probability distribution of the observable (\ref{apprqs1})\ has the
density%
\begin{equation}
p_{\alpha }(x)=\Tr \,\rho _{\alpha }(x)  \label{pd1}
\end{equation}%
and the posterior states are
\begin{equation}
\hat{\rho}_{\alpha }(x)=\rho _{\alpha }(x)/p_{\alpha }(x).  \label{post1}
\end{equation}

\begin{lemma}
\label{l1} \emph{The probability density (\ref{pd1}) is Gaussian:
\begin{equation}
p_{\alpha }(x)=\frac{1}{\sqrt{\left( 2\pi \right) ^{s}\det \left( \alpha
_{qq}+\beta \right) }}\exp \left[ -\frac{1}{2}x^{t}\left( \alpha _{qq}+\beta
\right) ^{-1}x\right] ,  \label{pd2}
\end{equation}%
and the posterior states (\ref{post1}) are Gaussian
\begin{equation}
\hat{\rho}_{\alpha }(x)=D(K_{q}x,K_{p}x)\rho _{\tilde{\alpha}%
}D(K_{q}x,K_{p}x)^{\ast },  \label{mean}
\end{equation}%
where
\begin{equation*}
K_{q}=\alpha _{qq}\left( \alpha _{qq}+\beta \right)
^{-1},\quad K_{p}=\alpha _{pq}\left( \alpha _{qq}+\beta \right) ^{-1} ,
\end{equation*}
 and $\rho_{\hat{\alpha}}$ is centered Gaussian state with the covariance matrix%
\begin{equation*}
\hat{\alpha}=\left[
\begin{array}{cc}
\hat{\alpha}_{qq} & \hat{\alpha}_{qp} \\
\hat{\alpha}_{pq} & \hat{\alpha}_{pp}%
\end{array}%
\right]
\end{equation*}%
with}
\begin{eqnarray}
\hat{\alpha}_{qq} &=&\left( \alpha _{qq}^{-1}+\beta ^{-1}\right) ^{-1},\quad
\notag \\
\hat{\alpha}_{pp} &=&\left( \alpha _{pp}-\alpha _{pq}(\alpha _{qq}+\beta
)^{-1}\alpha _{qp}\right) +\frac{1}{4}\beta ^{-1},  \label{corr} \\
\hat{\alpha}_{qp} &=&\left( \alpha _{qq}^{-1}+\beta ^{-1}\right) ^{-1}\alpha
_{qq}^{-1}\alpha _{qp}.  \notag
\end{eqnarray}
\end{lemma}

\textit{Proof}. From the quantum characteristic function of $\rho _{\alpha }$
we have%
\begin{equation*}
\Tr \,\rho _{\alpha }\exp \left( iu^{t}q\right) =\exp \left(
-\frac{1}{2}u^{t}\alpha _{qq} u\right) ,
\end{equation*}%
hence the diagonal value of the kernel of $\rho _{\alpha }$ in the position
representation is the inverse Fourier transform%
\begin{equation*}
\langle \xi |\rho _{\alpha }|\xi \rangle =\frac{1}{\sqrt{\left( 2\pi \right)
^{s}\det \alpha _{qq}}}\exp \left[ -\frac{1}{2}\xi ^{t}\alpha _{qq}^{-1}\xi %
\right] ,
\end{equation*}%
and
\begin{equation*}
p_{\alpha }(x)=\Tr \,\rho _{\alpha }(x)=\int \langle \xi
|\rho _{\alpha }|\xi \rangle \exp \left[ -\frac{1}{2}\left( \xi -x\right)
^{t}\beta ^{-1}\left( \xi -x\right) \right] \frac{d^{s}\xi }{\sqrt{\left(
2\pi \right) ^{s}\det \beta }}
\end{equation*}%
is the convolution of the two Gaussian probability densities giving the
right-hand side of (\ref{pd2}).

The posterior state (\ref{post1}) is Gaussian since it has the Gaussian
kernel in the Schr\"{o}dinger representation%
\begin{eqnarray*}
\langle \xi |\hat{\rho}_{\alpha }(x)|\xi ^{\prime }\rangle &=&\frac{1}{\sqrt{%
\left( 2\pi \right) ^{s}\det \beta }} \\
&\times &\exp \left[ -\frac{1}{4}\left( \xi -x\right) ^{t}\beta ^{-1}\left(
\xi -x\right) -\frac{1}{4}\left( \xi ^{\prime }-x\right) ^{t}\beta
^{-1}\left( \xi ^{\prime }-x\right) \right] \\
&\times &\langle \xi |\rho _{\alpha }|\xi ^{\prime }\rangle /p_{\alpha }(x),
\end{eqnarray*}%
where $\langle \xi |\rho _{\alpha }|\xi ^{\prime }\rangle $ is given by (\ref%
{mm0}) and $p_{\alpha }(x)$ -- by (\ref{pd2}). Substituting (\ref{mm0}%
) and (\ref{pd2}), we obtain%
\begin{equation*}
\langle \xi |\hat{\rho}_{\alpha }(x)|\xi ^{\prime }\rangle =\sqrt{\frac{\det
\left( \alpha _{qq}+\beta \right) }{\left( 2\pi \right) ^{s}\det \alpha
_{qq}\det \beta }}\exp \left[ \hat{\mu}\left( x;\xi ,\xi ^{\prime }\right) -%
\frac{1}{2}\hat{\theta}\left( \xi ,\xi ^{\prime }\right) \right] ,
\end{equation*}%
where%
\begin{equation*}
\hat{\mu}\left( x;\xi ,\xi ^{\prime }\right) =\frac{1}{2}x^{t}\beta ^{-1}\xi
+\frac{1}{2}x^{t}\beta ^{-1}\xi ^{\prime }-\frac{1}{2}x^{t}\beta ^{-1}x+%
\frac{1}{2}x^{t}\left( \alpha _{qq}+\beta \right) ^{-1}x,
\end{equation*}%
\begin{equation*}
\hat{\theta}\left( \xi ,\xi ^{\prime }\right) =\frac{1}{2}\xi ^{t}\beta
^{-1}\xi +\frac{1}{2}\left( \xi ^{\prime }\right) ^{t}\beta ^{-1}\xi
^{\prime }+\theta \left( \xi ,\xi ^{\prime }\right)
\end{equation*}%
and $\theta \left( \xi ,\xi ^{\prime }\right) $ is given by (\ref{theta}).
Let us first consider the terms independent of $\ x.$ We have%
\begin{eqnarray*}
\hat{\theta}\left( \xi ,\xi ^{\prime }\right) &=&\frac{1}{4}\left( \xi
^{\prime }+\xi \right) ^{t}\beta ^{-1}\left( \xi ^{\prime }+\xi \right) +%
\frac{1}{4}\left( \xi ^{\prime }-\xi \right) ^{t}\beta ^{-1}\left( \xi
^{\prime }-\xi \right) +\theta \left( \xi ,\xi ^{\prime }\right) \\
&=&\frac{1}{4}\left( \xi ^{\prime }+\xi \right) ^{t}\hat{\alpha}%
_{qq}^{-1}\left( \xi ^{\prime }+\xi \right) +\left( \xi ^{\prime }-\xi
\right) ^{t}(\hat{\alpha}_{pp}-\hat{\alpha}_{pq}\hat{\alpha}_{qq}^{-1}\hat{%
\alpha}_{qp})\left( \xi ^{\prime }-\xi \right) \\
&+&i\left( \xi ^{\prime }+\xi \right) ^{t}\hat{\alpha}_{qq}^{-1}\hat{\alpha}%
_{qp}\left( \xi ^{\prime }-\xi \right) ,
\end{eqnarray*}%
where
\begin{eqnarray*}
\hat{\alpha}_{qq}^{-1} &=&\alpha _{qq}^{-1}+\beta ^{-1}, \\
\hat{\alpha}_{qq}^{-1}\hat{\alpha}_{qp} &=&\alpha _{qq}^{-1}\alpha _{qp}, \\
\hat{\alpha}_{pp}-\hat{\alpha}_{pq}\hat{\alpha}_{qq}^{-1}\hat{\alpha}_{qp}
&=&\left( \alpha _{pp}-\alpha _{pq}\alpha _{qq}^{-1}\alpha _{qp}\right) +%
\frac{1}{4}\beta ^{-1},
\end{eqnarray*}%
resulting in the elements of the posterior covariance matrix (\ref{corr}).

To find the posterior mean values $\hat{m}_{q},\hat{m}_{p}$, we have
\begin{eqnarray*}
\hat{\mu}\left( x;\xi ,\xi ^{\prime }\right) &=&\frac{1}{2}x^{t}\beta
^{-1}\left( \xi +\xi ^{\prime }\right) -\frac{1}{2}x^{t}\beta ^{-1}\left(
\alpha _{qq}^{-1}+\beta ^{-1}\right) ^{-1}\beta ^{-1}x \\
&=&\frac{1}{2}\left( K_{q}x\right) ^{t}\hat{\alpha}_{qq}^{-1}\left( \xi +\xi
^{\prime }\right) -\frac{1}{2}\left( K_{q}x\right) ^{t}\hat{\alpha}%
_{qq}^{-1}\left( K_{q}x\right) ,
\end{eqnarray*}%
where $K_{q}=\alpha _{qq}\left( \alpha _{qq}+\beta \right) ^{-1}.$ Comparing
with the corresponding terms under the exponent in (\ref{mm}) (see Appendix), we should
have for the posterior mean values$:$%
\begin{equation*}
\hat{m}_{q}=K_{q}x,\quad \hat{m}_{p}=\hat{\alpha}_{pq}\hat{\alpha}_{qq}^{-1}%
\hat{m}_{q},
\end{equation*}%
where the second relation follows from the fact that the second term in (\ref%
{mm}) is zero in our case. Thus $\hat{m}_{p}=K_{p}x,$ where $K_{p}=\alpha
_{pq}\left( \alpha _{qq}+\beta \right) ^{-1}.$ $\square $

Let $\mathfrak{S}(\alpha )$ be the set of all (not necessarily centered)
states $\rho $ with the covariance matrix $\alpha .$ We will study the
following entropic characteristic of the Gaussian measurement $M$
\begin{equation}
ER(M;\alpha )=\sup_{\rho \in \mathfrak{S}(\alpha )}ER(\rho ,M),
\label{cchi1}
\end{equation}%
which is strictly related to the entanglement-assisted capacity of $M$ (see
sec. \ref{s3}). Due to (\ref{zerom}), in (\ref{cchi1}) we can restrict to
centered states $\rho $ for which $\alpha $ coincides with the matrix of
second moments. We denote $$g(x)=(x+1)\log (x+1)-x\log x .$$
\begin{theorem}
\label{t3} \emph{The supremum in (\ref{cchi1}) is attained on the Gaussian
state $\rho _{\alpha }$ and it is equal to
\begin{equation}
ER(M;\alpha )=\frac{1}{2}\left[ \mathrm{Sp}\,g\left( \mathrm{abs}(\Delta
^{-1}\alpha )-\frac{I_{2s}}{2}\right) -\mathrm{Sp}\,g\left( \mathrm{abs}%
(\Delta ^{-1}\hat{\alpha})-\frac{I_{2s}}{2}\right) \right] ,  \label{cchi2}
\end{equation}%
where $\hat{\alpha}$ is given by (\ref{corr}), and }$\mathrm{abs}(\Delta
^{-1}\alpha )$ \emph{denotes the matrix with eigenvalues equal to modulus of
eigenvalues of } $\Delta ^{-1}\alpha $ \ \emph{and with the same eigenvectors%
$.$ }
\end{theorem}

\emph{Proof.} Let $\rho $ be a centered density operator from $\mathfrak{S}%
(\alpha )$, then denote $\rho (x)=V(x)\,\rho \,V(x)$ and $p(x)=\Tr %
\rho (x).$ Also introduce the channel $\mathcal{E}\,[\rho ]=\,\left\{
V(x)\,\rho \,V(x)\right\} $ with quantum input and hybrid \textit{%
classical-quantum} (cq)-output. For the Gaussian state  $\rho _{\alpha }$ we
have
\begin{eqnarray}
&&ER(\rho _{\alpha },M)-ER(\rho ,M)  \label{difer} \\
&=&\mathrm{H}(\rho \Vert \rho _{\alpha })+\Tr (\rho -\rho _{\alpha
})\log \rho_{\alpha }   \notag \\
&-&\mathrm{H}_{cq}(\mathcal{E}\,[\rho ]\Vert \mathcal{E}\,[\rho _{\alpha
}])\,\,+\mathrm{H}_{c}(p\Vert p_{\alpha })  \notag \\
&+&\int \Tr (\rho _{\alpha }(x)-\rho (x))\log \,\hat{\rho}_{\alpha
}(x)d^{s}x.  \notag
\end{eqnarray}%
Here
\begin{equation*}
\mathrm{H}_{c}(p\Vert p_{\alpha })=\int p(x)\log \left( \frac{p(x)}{%
p_{\alpha }(x)}\right) d^{s}x
\end{equation*}%
is the classical relative entropy between $p$, $p_{\alpha }$ and
\begin{equation*}
\mathrm{H}_{cq}(\mathcal{E}\,[\rho ]\Vert \mathcal{E}\,[\rho _{\alpha
}])=\int \Tr \,\rho (x)\,\left( \log \rho (x)-\log \rho
_{\alpha }(x)\right) d^{s}x,
\end{equation*}%
is the relative entropy of the cq-states (see Eq. (3) in \cite{BL-2}).

Monotonicity of the relative entropy for cq-states (\cite{BL-2}, theorem 1)
then implies
\begin{equation*}
\mathrm{H}_{cq}(\mathcal{E}\,[\rho ]\Vert \mathcal{E}\,[\rho _{\alpha
}])\leq \mathrm{H}(\rho \Vert \rho _{\alpha }),
\end{equation*}%
hence we have for the first three terms in the right-hand side of (\ref%
{difer})
\begin{equation}
\mathrm{H}(\rho \Vert \rho _{\alpha })-\mathrm{H}_{cq}(\mathcal{E}\,[\rho
]\Vert \mathcal{E}\,[\rho _{\alpha }])+\mathrm{H}_{c}(p\Vert p_{\alpha
})\geq 0.  \label{first}
\end{equation}

Without loss of generality we can assume that $\rho _{\alpha }$ is
non-degenerate so that $\log \rho _{\alpha }$ exists and it is a polynomial
in $q,p$ of the second order$.$ This follows from the exponential form of
the density operator (theorem 12.23 in \cite{QSCI}). Since the first and
second moments of the states $\rho $ and $\rho _{\alpha }$ coincide, we have
\begin{equation}
\Tr (\rho -\rho _{\alpha })\log \rho _{\alpha }=0.  \label{second}
\end{equation}%
It remains to show that also
\begin{equation}
\int \Tr (\rho _{\alpha }(x)-\rho (x))\log \,\hat{\rho}_{\alpha
}(x)\,d^{s}x=0.  \label{res}
\end{equation}

Substituting the posterior state (\ref{mean}) into the right-hand side of (%
\ref{res}), we obtain
\begin{eqnarray}
\int \Tr (\rho _{\alpha }(x)-\rho (x))\log \hat{\rho}_{\alpha
}(x)d^{s}x &=&\Tr \,\Phi _{M}\,[\rho _{\alpha }-\rho ]\,\log \rho _{%
\hat{\alpha}}  \label{res3} \\
&=&\Tr \,\,(\rho _{\alpha }-\rho )\,\Phi _{M}^{\ast }[\log \rho _{%
\hat{\alpha}}],  \label{res3a}
\end{eqnarray}%
where we have introduced the channel
\begin{equation}
\Phi _{M}[\sigma ]=\int D(K_q x, K_p x)^{\ast }V(x)\,\sigma \,V(x)D(K_q x, K_p x)d^{s}x.
\label{M-add}
\end{equation}%
The channel $\Phi _{M}$ is Gaussian. First, it is a c.p.t.p. map. Complete positivity is apparent from the structure of the
map (\ref{M-add}). Trace preservation follows from
\begin{equation*}
\Phi _{M}^{\ast }[I]=\int V(x)^{2}\,d^{s}x=\int M(\,d^{s}x)=I.
\end{equation*}%
A routine calculation shows that if $\sigma $ is a Gaussian state then $\Phi
_{M}[\sigma ]$ is again Gaussian. Then by result of \cite{dpgh}, $\Phi _{M}$
is a Gaussian channel.

Since $\rho _{\hat{\alpha}}$ is Gaussian state, $\log \rho _{\hat{\alpha}}$
is a polynomial in $q,p$ of the second order. The dual Gaussian channel $%
\Phi _{M}^{\ast }$ takes it into another polynomial in $q,p$ of the second
order (see sec. 12.4.2 of \cite{QSCI}). Since the first and second moments of the states $\rho $ and $\rho
_{\alpha }$ coincide, one obtains (\ref{res}). Then (\ref{difer}), (\ref%
{first}), (\ref{second}) and (\ref{res}) imply%
\begin{equation*}
ER(\rho _{\alpha },M)-ER(\rho ,M)\geq 0
\end{equation*}%
for arbitrary $\rho \in \mathfrak{S}(\alpha )$ proving that the supremum (%
\ref{cchi1}) is attained on the Gaussian state $\rho _{\alpha }$.

Taking into account the formula (\ref{mean}) for the posterior states, \ we
obtain
\begin{eqnarray*}
ER(M;\alpha ) &=&ER(\rho _{\alpha },M) \\
&=&\mathrm{H}(\rho _{\alpha })-\int_{\Omega }p_{\alpha }(x)\,\mathrm{H}(\hat{%
\rho}_{\alpha }(x))d^{s}x \\
&=&\mathrm{H}(\rho _{\alpha })-\int p_{\alpha }(x)\mathrm{H}(D(K_q x, K_p x)^{\ast
}\rho _{\hat{\alpha}}D(K_q x, K_p x))d^{s}x \\
&=&\mathrm{H}(\rho _{\alpha })-\mathrm{H}(\rho _{\hat{\alpha}}).
\end{eqnarray*}%
Then (\ref{cchi2}) follows by applying the formula for the entropy of a
Gaussian state%
\begin{equation}
\mathrm{H}(\rho _{\alpha })=\frac{1}{2}\mathrm{Sp}\,g\left( \mathrm{abs}%
(\Delta ^{-1}\alpha )-\frac{I_{2s}}{2}\right) ,  \label{entro}
\end{equation}%
given in \cite{QSCI}, Eq. (12.110). $\square $

\section{The entanglement-assisted capacity}
\label{s3}

Consider a quadratic Hamiltonian of the form%
\begin{equation*}
H=\sum_{j,k=1}^{s}\left( q^{t}\epsilon _{qq}q+q^{t}\epsilon
_{qp}p+p^{t}\epsilon _{pq}q+p^{t}\epsilon _{pp}p\right) ,
\end{equation*}%
where
\begin{equation*}
\epsilon =\left[
\begin{array}{cc}
\epsilon _{qq} & \epsilon _{qp} \\
\epsilon _{pq} & \epsilon _{pp}%
\end{array}%
\right]
\end{equation*}%
is a real symmetric positive definite energy matrix. The importance of the
quantity (\ref{cchi1}) is that it underlies the energy-constrained
entanglement-assisted capacity, given by the formula (see e.g. \cite{H3})
\begin{equation*}
C_{ea}(M,H,E)=\max_{\rho :~\Tr \rho H\leq E}ER(\rho ,M),
\end{equation*}%
where $E$ is the energy level. For any state $\rho \in \mathfrak{S}(\alpha )$
the mean energy\footnote{%
We denote $\mathrm{Sp}$ trace of matrices as distinct from the trace of
operators in $\mathcal{H}$.}
\begin{equation*}
\Tr \rho H=\mathrm{Sp}\,\epsilon \alpha _{2}\geq \mathrm{Sp}%
\,\epsilon \alpha ,
\end{equation*}%
where $\alpha _{2}$ is the matrix of second moments of $\rho ,$ with the
equality attained for centered states, therefore
\begin{equation}
C_{ea}(M;H,E)=\max_{\alpha :\mathrm{Sp}\,\epsilon \alpha \leq E}ER(M;\alpha
).  \label{ceag}
\end{equation}
It is easy to check that%
\begin{equation*}
\sup_{\alpha }ER(M;\alpha )=+\infty ,
\end{equation*}%
therefore theorem 1 of \cite{H3} applies showing that the maximum in (\ref{ceag}) is
attained on $\alpha $ satisfying $\mathrm{Sp}\,\epsilon \alpha =E.$

In the case of the oscillator-type Hamiltonian%
\begin{equation*}
H=\sum_{j,k=1}^{s}\left( q^{t}\epsilon _{qq}q+p^{t}\epsilon _{pp}p\right) ,
\end{equation*}%
the energy constraint is
\begin{equation}
\mathrm{Sp}\,\epsilon _{qq}\alpha _{qq}+\mathrm{Sp}\,\epsilon _{pp}\alpha
_{pp}\leq E.  \label{ed}
\end{equation}%
Then the maximization in (\ref{ceag}) can be taken over only block-diagonal
covariance matrices $\alpha \ ($satisying (\ref{ed}) with equality). To see
this consider the transformation $T:q_j\rightarrow q_j,\,p_j\rightarrow -p_j$ which
changes the sign of the commutators between $q_j$ and $p_j$ and which is
implemented by anti-unitary operator of complex conjugation in the Schr\"odinger representation. If $\rho $
is a centered state with the covariance matrix \footnote{%
We denote by $I_{s}$ the unit $s\times s-$matrix.}
\begin{equation*}
\alpha =\left[
\begin{array}{cc}
\alpha _{qq} & \alpha _{qp} \\
\alpha _{pq} & \alpha _{pp}%
\end{array}%
\right] \geq \pm \frac{i}{2}\left[
\begin{array}{cc}
0 & I_{s} \\
I_{s} & 0%
\end{array}%
\right] ,
\end{equation*}%
then $T[\rho ]$ has the covariance matrix
\begin{equation*}
\tilde{\alpha} =\left[
\begin{array}{cc}
\alpha _{qq} & -\alpha _{qp} \\
-\alpha _{pq} & \alpha _{pp}%
\end{array}%
\right] \geq \mp \frac{i}{2}\left[
\begin{array}{cc}
0 & I_{s} \\
I_{s} & 0%
\end{array}%
\right] ,
\end{equation*}%
because it is the covariance matrix for $q_j,\,-p_j.$ The mixture $\frac{1}{2}%
\left( \rho +T[\rho ]\right) $ has the covariance matrix
\begin{equation*}
\left[
\begin{array}{cc}
\alpha _{qq} & 0 \\
0 & \alpha _{pp}%
\end{array}%
\right] =\frac{1}{2}\left( \alpha +\tilde{\alpha}\right) \geq \pm \frac{i}{2}%
\left[
\begin{array}{cc}
0 & I_{s} \\
I_{s} & 0%
\end{array}%
\right] ,
\end{equation*}%
while
\begin{equation*}
ER(\frac{1}{2}\left( \rho +T[\rho ]\right) ,M)\geq \frac{1}{2}\left[ ER(\rho
,M)+ER(T[\rho ],M)\right] =ER(\rho ,M)
\end{equation*}%
by the concavity of the entropy reduction and its invariance under the
transformation $T$.

\section{The case of one mode}
\label{s4}

\begin{figure}[t]
\center{\includegraphics{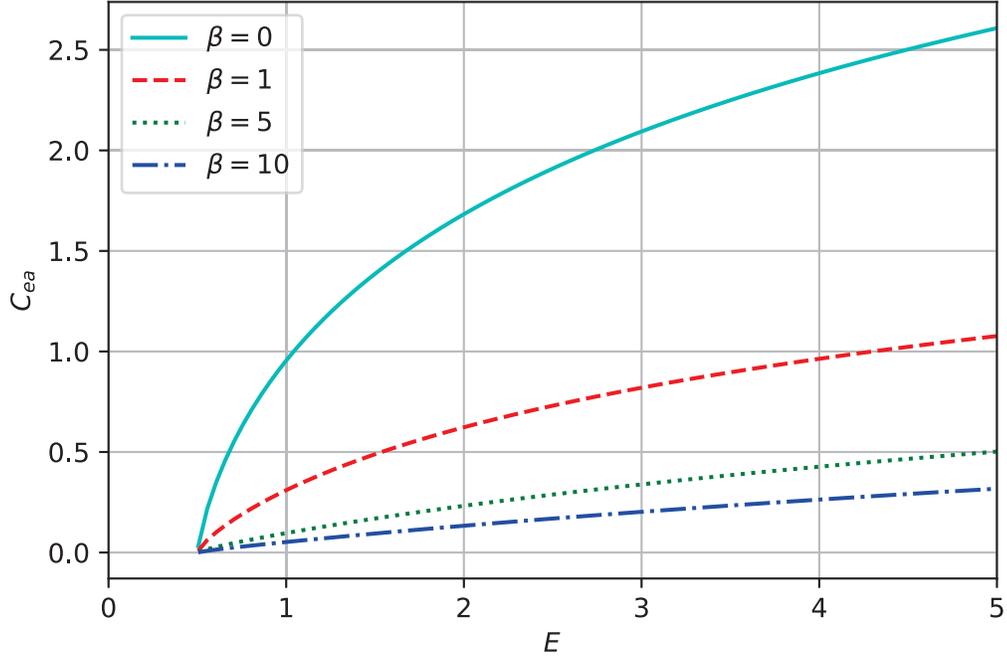}}
\caption{(color online) Plot of the assisted capacity $C_{ea}(E)$ for $\beta=0, 1, 5, 10$.}
\label{Fig1}
\end{figure}

Consider the approximate measurement of position for one bosonic mode $q,p$.
The corresponding POVM is
\begin{equation}
M(dx)=\exp \left[ -\frac{\left( q-x\right) ^{2}}{2\beta }\right] \frac{dx}{%
\sqrt{2\pi \beta }}=D(x)\mathrm{e}^{-q^{2}/2\beta }D(x)^{\ast }\frac{dx}{%
\sqrt{2\pi \beta }},  \label{apprq}
\end{equation}%
where $\beta >0.$ Probability distribution for the Gaussian state $\rho
_{\alpha }$%
\begin{equation}
\Tr \,\rho _{\alpha }M(dx)=\exp \left[ -\frac{x^{2}}{2\left(
\alpha _{qq}+\beta \right) }\right] \frac{dx}{\sqrt{2\pi \left( \alpha
_{qq}+\beta \right) }}=p_{\alpha }(x)\,dx,  \label{pd}
\end{equation}%
with the covariance $2\times 2-$matrix (\ref{cm}), where $\alpha
_{qq},\alpha _{qp},\alpha _{pp}$ are real numbers satisfying
\begin{equation}
\alpha _{qq}\alpha _{pp}-\alpha _{qp}^{2}\geq 1/4.  \label{ur}
\end{equation}%
The formula (\ref{entro}) for the entropy amounts to (see Example 12.25 in
\cite{QSCI})%
\begin{equation*}
\mathrm{H}(\rho _{\alpha })=g\left( \sqrt{\alpha _{qq}\alpha _{pp}-\alpha
_{qp}^{2}}-\frac{1}{2}\right) .
\end{equation*}

According to eq. (\ref{corr}) posterior states have the covariance $2\times 2$-matrix $%
\hat{\alpha}$ with the elements%
\begin{equation*}
\hat{\alpha}_{qq}=\frac{\alpha _{qq}\beta }{\alpha _{qq}+\beta },\quad \hat{%
\alpha}_{pp}=\left( \alpha _{pp}-\frac{\alpha _{qp}^{2}}{\alpha _{qq}+\beta }%
\right) +1/\left( 4\beta \right) ,\quad \hat{\alpha}_{qp}=\frac{\alpha
_{qp}\beta }{\alpha _{qq}+\beta }.
\end{equation*}%
Then theorem \ref{t3} gives%
\begin{eqnarray}
ER(M;\alpha ) &=&g\left( \sqrt{\alpha _{qq}\alpha _{pp}-\alpha _{qp}^{2}}-%
\frac{1}{2}\right) -g\left( \sqrt{\hat{\alpha}_{qq}\hat{\alpha}_{pp}-\hat{%
\alpha}_{qp}^{2}}-\frac{1}{2}\right)  \notag \\
&=&g\left( \sqrt{\alpha _{qq}\alpha _{pp}-\alpha _{qp}^{2}}-\frac{1}{2}%
\right)  \notag \\
&-&g\left( \sqrt{\frac{\alpha _{qq}(\beta \alpha _{pp}+1/4)-\beta \alpha
_{qp}^{2}}{\alpha _{qq}+\beta }}-\frac{1}{2}\right) .  \label{erm}
\end{eqnarray}

For the oscillator Hamiltonian $H=\frac{1}{2}\left( q^{2}+p^{2}\right) ,$
the energy constraint has the form $\alpha _{qq}+\alpha _{pp}\leq 2E$ and
the entanglement-assisted capacity (\ref{ceag}) is equal to%
\begin{equation}
C_{ea}(M;H,E)=  \label{ceab}
\end{equation}%
\begin{equation*}
=\max \left[ g\left( \sqrt{\alpha
_{qq}\,\alpha _{pp}}-\frac{1}{2}\right) -g\left( \sqrt{\frac{\alpha
_{qq}(\beta \alpha _{pp}+1/4)}{\alpha _{qq}+\beta }}-\frac{1}{2}\right) %
\right] ,
\end{equation*}
where the maximum over positive $\alpha _{qq},\alpha _{pp}$ satisfying $%
\alpha _{qq}+\alpha _{pp}= 2E,$ $\alpha _{qq}\alpha _{pp}\geq 1/4$, and $\alpha_{qp}$ is taken to be $0$ (see Fig. \ref{Fig1}).

Let us now consider the limit $\beta \rightarrow 0$ corresponding to the
exact measurement of position, where
\begin{equation}
M_{ex}(dx)=|x\rangle \langle x|dx  \label{exactq}
\end{equation}%
is the spectral measure of the operator $q$. Notice that strictly speaking
it is not of the form (\ref{den}), i.e. it does not have a bounded
operator-valued density. In the limit $\beta \rightarrow 0$ we have%
\begin{equation}
ER(M;\alpha )\rightarrow g\left( \sqrt{\alpha _{qq}\,\alpha _{pp}}-\frac{1}{2%
}\right) .  \label{ER2}
\end{equation}%
We obtain the entanglement-assisted capacity
\begin{equation}
C_{ea}\left( M_{ex},H,E\right) =\max_{}g\left( \sqrt{\alpha _{qq}\,\alpha
_{pp}}-\frac{1}{2}\right) ,  \label{cerm}
\end{equation}%
where the maximum over positive $\alpha _{qq},\alpha _{pp}$ satisfying $%
\alpha _{qq}+\alpha _{pp}= 2E,$ $\alpha _{qq}\alpha _{pp}\geq 1/4.$ By
the concavity of $g(x)$, this maximum is attained for $\alpha _{qq}=\alpha
_{pp}=E$ and is equal to%
\begin{equation}
C_{ea}\left( M_{ex},H,E\right) =g(E-1/2).  \label{ceaex}
\end{equation}

In \cite{H2} the POVM (\ref{exactq}) was considered as associated with the
instrument
\begin{equation*}
\mathcal{M}_{ex}[\rho ](dx)=|e\rangle \langle x|\rho |x\rangle \langle
e|\,dx,
\end{equation*}%
where $|e\rangle $ is an arbitrary unit vector. Notice that here $%
V(x)=|e\rangle \langle x|$ is an unbounded and even non-closable operator.
Then the posterior state $|e\rangle \langle e|$ is pure and has zero
entropy, resulting in $ER(M_{ex};\rho )=\mathrm{H}(\rho ).$ Thus
\begin{equation}
ER(M_{ex};\alpha )=\mathrm{H}(\rho _{\alpha })=g\left( \sqrt{\alpha
_{qq}\,\alpha _{pp}}-\frac{1}{2}\right) ,  \label{ER1}
\end{equation}%
which agrees with (\ref{ER2}) and leads again to the capacity (\ref{ceaex}).

Let us consider another realization of the exact measurement -- the squeezed
joint measurement of $q,p,$ described by POVM
\begin{equation*}
M_{sq}(dx\,dy)=|x,y\rangle _{\beta }\langle x,y|\frac{dx\,dy}{2\pi }%
=D(x,y)\rho _{sq,\beta }\,D(x,y)^{\ast }\frac{dx\,dy}{2\pi },
\end{equation*}%
where $\rho _{sq,\beta }$ is the squeezed vacuum with the covariance matrix $%
\left[
\begin{array}{cc}
\beta  & 0 \\
0 & 1/\left( 4\beta \right)
\end{array}%
\right] .$ Then
\begin{equation}
\int_{y}M_{sq}(dx\,dy)=M(dx).  \label{ave}
\end{equation}%
To see this it is sufficient to show that
\begin{equation}
\int D(x,y)\rho _{sq,\beta }D(x,y)^{\ast }\frac{\,dy}{2\pi }=\frac{1}{\sqrt{%
2\pi \beta }}\exp \left[ -\frac{\left( q-x\right) ^{2}}{2\beta }\right] .
\label{int}
\end{equation}%
Indeed, in the position representation%
\begin{eqnarray*}
&&\langle \xi |\int D(x,y)\rho _{sq,\beta }D(x,y)^{\ast }\frac{\,dy}{2\pi }%
|\xi ^{\prime }\rangle  \\
&=&\delta (\xi -\xi ^{\prime })\langle \xi -x|\rho _{sq,\beta }|\xi
-x\rangle  \\
&=&\langle \xi |\xi ^{\prime }\rangle \langle \xi -x|\rho _{sq,\beta }|\xi
-x\rangle .
\end{eqnarray*}%
The classical characteristic function is
\begin{equation*}
\int \langle \xi |\rho _{sq,\beta }|\xi \rangle \mathrm{e}^{i\xi \lambda
}d\xi =\Tr \,\rho _{sq,\beta }\mathrm{e}^{iq\lambda }=\exp \left( -%
\frac{1}{2}\beta \lambda ^{2}\right),
\end{equation*}%
whence%
\begin{equation*}
\langle \xi -x|\rho _{sq,\beta }|\xi -x\rangle =\frac{1}{\sqrt{2\pi \beta }}%
\exp \left[ -\frac{\left( \xi -x\right) ^{2}}{2\beta }\right] .
\end{equation*}%
which proves (\ref{int}).

Similarly to (\ref{ER1}) we have%
\begin{equation*}
ER(M_{sq};\alpha )=\mathrm{H}(\rho _{\alpha })=g\left( \sqrt{\alpha
_{qq}\,\alpha _{pp}}-\frac{1}{2}\right) .
\end{equation*}%
The difference between this expression and (\ref{erm}) reflects the loss of
information due to the averaging (\ref{ave}) over the values of the momentum.

\textbf{Acknowledgment}. The work was supported by the grant of Russian
Science Foundation (project No 19-11-00086).

\section{Appendix}

\emph{The kernel of a Gaussian density operator $\rho _{m,\alpha }$ with the mean $%
m_{q},m_{p}$ and the covariance matrix (\ref{cm}) in the Schr\"{o}dinger
representation has the form}
\begin{equation}
\langle \xi |\rho _{\alpha }|\xi ^{\prime }\rangle =\frac{1}{\sqrt{\left(
2\pi \right) ^{s}\det \alpha _{qq}}}\exp \left[ \mu \left( m;\xi ,\xi
^{\prime }\right) -\frac{1}{2}\theta \left( \xi ,\xi ^{\prime }\right) %
\right] ,  \label{mm}
\end{equation}%
\emph{where}%
\begin{equation}
\mu \left( m;\xi ,\xi ^{\prime }\right) =\frac{1}{2}m_{q}^{t}\alpha
_{qq}^{-1}\left( \xi ^{\prime }+\xi \right) -i\left( m_{p}-\alpha
_{pq}\alpha _{qq}^{-1}m_{q}\right) ^{t}\left( \xi ^{\prime }-\xi \right) -%
\frac{1}{2}m_{q}^{t}\alpha _{qq}^{-1}m_{q},  \label{mu}
\end{equation}%
\begin{eqnarray}
\theta \left( \xi ,\xi ^{\prime }\right)  &=&\frac{1}{4}\left( \xi ^{\prime
}+\xi \right) ^{t}\alpha _{qq}^{-1}\left( \xi ^{\prime }+\xi \right) +\left(
\xi ^{\prime }-\xi \right) ^{t}\left( \alpha _{pp}-\alpha _{pq}\alpha
_{qq}^{-1}\alpha _{qp}\right) \left( \xi ^{\prime }-\xi \right)   \notag \\
&+&i\left( \xi ^{\prime }+\xi \right) ^{t}\alpha _{qq}^{-1}\alpha
_{qp}\left( \xi ^{\prime }-\xi \right) .  \label{teta}
\end{eqnarray}

\emph{Proof.} The quantum characteristic function of $\rho _{m,\alpha }$ is
\begin{equation*}
\varphi (x,y)=\Tr \rho _{m,\alpha }W(x,y)=\exp [im(x,y)-\frac{1}{2}%
\alpha (x,y)],
\end{equation*}%
where
\begin{equation*}
\begin{aligned} m(x,y) &= m_q x + m_p y ,\\ \alpha(x,y) &= x^t \alpha_{qq} x
+ x^t \alpha_{qp} y + y^t \alpha_{pq} x + y^t \alpha_{pp} y \end{aligned}
\end{equation*}%
and $W(x,y)=e^{ix^{t}y/2}e^{ix^{t}q}e^{iy^{t}p}$ is the Weyl operator. The
kernel of the Weyl operator is
\begin{equation*}
\bra{\xi}W(x,y)\ket{\xi'}=\exp \left[ {i}\left( \frac{\xi +\xi ^{\prime }}{2}%
\right) ^{t}x\right] \delta (y-(\xi ^{\prime }-\xi ))=\exp \left(
iu^{t}x\right) \delta (y-v).
\end{equation*}%
Here we introduced the variables $u=\frac{\xi +\xi ^{\prime }}{2}$, $v=\xi ^{\prime }-\xi $.
Using the inversion formula quantum Fourier transform, we  readily compute
the kernel of the Gaussian state:
\begin{multline*}
\bra{\xi}\rho _{m,\alpha }\ket{\xi'}=\int \frac{d^{s}xd^{s}y}{(2\pi )^{s}}%
\varphi (x,y)\bra{\xi}W(-x,-y)\ket{\xi'}= \\
=\int \frac{d^{s}x}{(2\pi )^{s}}\exp \left[ -iu^{t}x+im(x,-v)-\frac{1}{2}%
\alpha (x,-v)\right] ,
\end{multline*}%
the expression under the exponent expands as
\begin{equation*}
-im_{p}^{t}v-\frac{1}{2}v^{t}\alpha _{pp}v-i(u^{t}-m_{q}^{t}+iv^{t}\alpha
_{pq})x-\frac{1}{2}x^{t}\alpha _{qq}x.
\end{equation*}%
Taking the $s$-dimensional Gaussian integral yields
\begin{multline*}
\bra{\xi}\rho _{\alpha }\ket{\xi'}=\frac{e^{-im_{p}^{t}v-\frac{1}{2}%
v^{t}\alpha _{pp}v}}{\sqrt{(2\pi )^{s}\det {\alpha _{qq}}}}\exp \left[ -%
\frac{1}{2}(u-m_{q}+i\alpha _{qp}v)^{t}\alpha _{qq}^{-1}(u-m_{q}+i\alpha
_{qp}v)\right] = \\
=\frac{1}{\sqrt{(2\pi )^{s}\det {\alpha _{qq}}}}\exp \left[ \mu (m;\xi ,\xi
^{\prime })-\frac{1}{2}\theta (\xi ,\xi ^{\prime })\right] ,
\end{multline*}%
where $\mu (m;\xi ,\xi ^{\prime })$ and  $\theta (\xi ,\xi ^{\prime })$ are
given by (\ref{mu}), (\ref{teta}). $\square $%

\end{document}